\documentclass[secnumarabic,aps,prl,reprint,groupedaddress]{revtex4-1}
\usepackage{verbatim}
\usepackage[english]{babel}
\usepackage{graphicx}
\usepackage{color}
\usepackage{amsmath}
\usepackage{soul}
\usepackage{tabularx}

\begin{document}
	
\title{Transport and relaxation of current-generated nonequilibrium phonons from nonlocal electronic measurements}
\author{Guanxiong Chen}
\author{Sergei Urazhdin}
\affiliation{Department of Physics, Emory University, Atlanta, GA, USA.}
	
\begin{abstract}

We study phonons generated by current in a Pt nanowire, by measuring resistance of another nanowire separated from the first one by an insulating spacer. For thin spacers, the resistance varies almost linearly with current at cryogenic temperatures, while an additional quadratic contribution emerges for thicker spacers. These observations suggest a non-thermal distribution of current-generated phonons that relaxes via strongly nonlinear dynamical processes rather than few-phonon scattering. Our results provide insight into the  nonequilibrium phonon dynamics at nanoscale, which may facilitate efficient heat management in electronic nanodevices.

\end{abstract}
	
\maketitle

The rapidly increasing speed and functionalities of modern electronic devices are enabled by the increasing complexity and density of integrated circuits (ICs) facilitated by the continued downscaling of circuit elements~\cite{noh2007downscaling}. The characteristic dimensions of transistors are reaching scales below $10$~nm~\cite{martinez2008silicon,desai2016mos2,ilatikhameneh2016saving}, approaching the fundamental limits of semiconductor technology~\cite{mamaluy2015fundamental}. Further density increases based, for example, on 3D IC architectures~\cite{wong2007monolithic,franzon2008design,knechtel2017large} will require new approaches to efficient dissipation of Joule heat at nanoscale~\cite{sekar20083d,wang20183d,xie2011electrical}.

Dissipation of current-generated heat is traditionally analyzed in terms of Fourier's law of diffusive heat flow. However, it is now recognized that diffusive approximation starts to break down in modern transistors whose dimensions are comparable to the phonon mean free path (MFP)~\cite{wang2016enhanced,escobar2006multi,narumanchi2003simulation,PhysRevB.98.085427}. Thus, optimizing heat dissipation at nanoscale is not equivalent to maximizing thermal conductivity~\cite{huang2015length,anufriev2017heat,regner2013broadband}.

Current-generated phonons are commonly assumed to form a quasi-equilibrium distribution characterized by local effective temperature~\cite{temperature}, as embodied by the interchangeable use of terms ''Joule dissipation" and ''Joule heating"~\cite{ghai2005novel,nabovati2011lattice}. The former describes the electric energy deposited in the material, while the latter implies that phonons generated by current form a thermal distribution and can be described as heat. However, recent observations of linear dependence of resistance on current in metallic nanostructures at cryogenic temperatures indicated a non-equilibrium phonon distribution that cannot be characterized by an effective temperature~\cite{chen2020observation}. Similar conclusions were reached in the studies of electrically biased carbon nanotubes~\cite{PhysRevLett.84.2941,PhysRevB.76.085433}
and the analysis of current-driven superconductivity suppression~\cite{ritter2021role}. These findings suggest the possibility to optimize heat dissipation, for instance, by utilizing a relatively small flux of non-thermalized high-energy phonons. However, it is presently not known whether non-thermal phonon distributions can be transported over substantial distances suitable for IC applications.

\begin{figure}
	\includegraphics[width=\columnwidth]{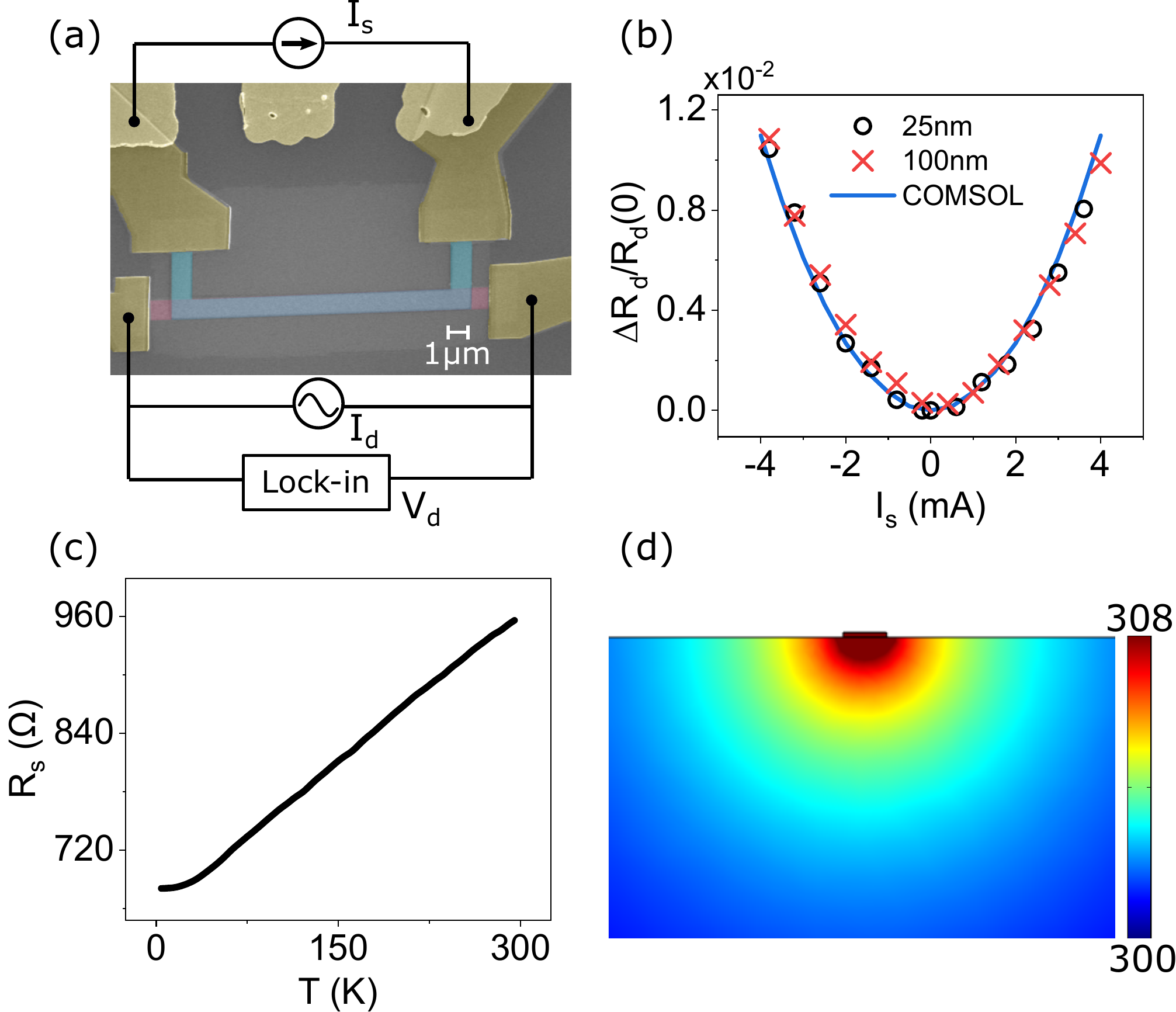}
	\caption{\label{fig:RT}(a) Pseudocolor SEM image of one of the studied samples and measurement setup. (b) $\Delta R_d/R_d(0)$ vs $I_s$ for $d=25$~nm (circles) and $100$~nm (crosses), at $T=300$~K. Here, $\Delta R_d=R_d(I_s)-R_d(0)$. Curve: COMSOL simulation for $d=100$~nm. The curve is well-approximated by a parabola. The result for $d=25$~nm [not shown] is nearly identical. (c) $R_s$ vs $T$ at $I_s=0$. The temperature dependence of $R_d$ is similar. (d) Crossection of the temperature distribution calculated for $d=100$~nm, at $I_s=4$~mA and $T=300$~K.}
\end{figure}

Here, we present nonlocal electronic measurements utilizing a phonon-detecting nanowire separated from the phonon-generating wire by an electrically insulating spacer. The separation between phonon generation and detection allows us to confirm non-thermal distribution of current-generated phonons. Our approach also allows us to characterize inelastic phonon scattering, and to elucidate its mechanisms.

Our samples were fabricated by e-beam lithography and sputtering. They consisted of two $7$~nm-thick, $1$~$\mu$m-wide and approximately $18$~$\mu$m-long Pt nanowires fabricated on top of one another on undoped Si substrates, and contacted by Cu electrodes [Fig.~\ref{fig:RT}(a)]. Their large length ensured that phonon escape into the electrodes was negligible. The wires were separated by an insulating amorphous SiO$_2$ spacer whose thickness $d$ was varied between $5$ and $100$~nm.  To generate phonons, dc current $I_s$ was applied to the top wire. The resistance $R_d$ of the bottom wire used as a phonon detector was simultaneously measured using lock-in detection with a small ac current $I_d=10$ $\mu$A applied to this wire. Both wires were metallic with resistance of about 800~$ \Omega $, with negligible contribution from contact resistance, as verified by separate 4-probe measurements. The resistance between the wires was at least 25~M$\Omega$ at cryogenic temperatures.

Figure~\ref{fig:RT}(b) shows $R_d(I_s)$ for two spacer thicknesses, $d=25$~nm and $100$~nm, at temperature $T=300$~K~\cite{highT}. To account for minor geometric differences leading to slightly different resistances, the data are offset and normalized by $R_d(0)$. The two datasets closely follow the same  quadratic dependence. This result is consistent with Joule heating and Fourier's law of heat diffusion. Indeed, the rate of Joule dissipation per unit source wire area is $w=d_{Pt}\rho j_s^2$, where $d_{Pt}$ is its thickness, $j_s$ is the current density, and $\rho$ is resistivity. For the studied thin-film wires, the latter is dominated by scattering on the surfaces and impurities, as evidenced by its weak dependence on $T$ [Fig.~\ref{fig:RT}(c)], so to the lowest order the variation of $\rho$ with  $j_s$ can be neglected.

The dissipated energy produces a heat flux $q=w$ flowing through the spacer and the sensing layer into the substrate and spreading over the characteristic depth $z_{c}$ defined by the wire width, as illustrated by the COMSOL simulation, Fig.~\ref{fig:RT}(d). In the 1d approximation, this can be modeled by a heat sink with the temperature $T$ at the depth $z_{c}$ below the sensing Pt wire. According to the Fourier's law, $q=-\kappa\nabla  T$, where $\kappa$ is the thermal conductivity of the substrate. We infer that the temperature of the detector wire $T_d=T+d_{Pt}\rho {j_s}^2/\kappa$ does not depend on the properties of the spacer, and is quadratic in $I_s$. Above the Bloch–Grüneisen temperature $\Theta_R\approx 50$~K~\cite{kittel1976introduction}, the resistance $R_d$ of the detector wire is linear in $T$ [Fig.~\ref{fig:RT}(c)], so $R_d$ is expected to be quadratic in $I_s$, in agreement with the data and the COMSOL simulations based on the Fourier's law [curve Fig.~\ref{fig:RT}(b)].

\begin{figure}
	\includegraphics[width=\columnwidth]{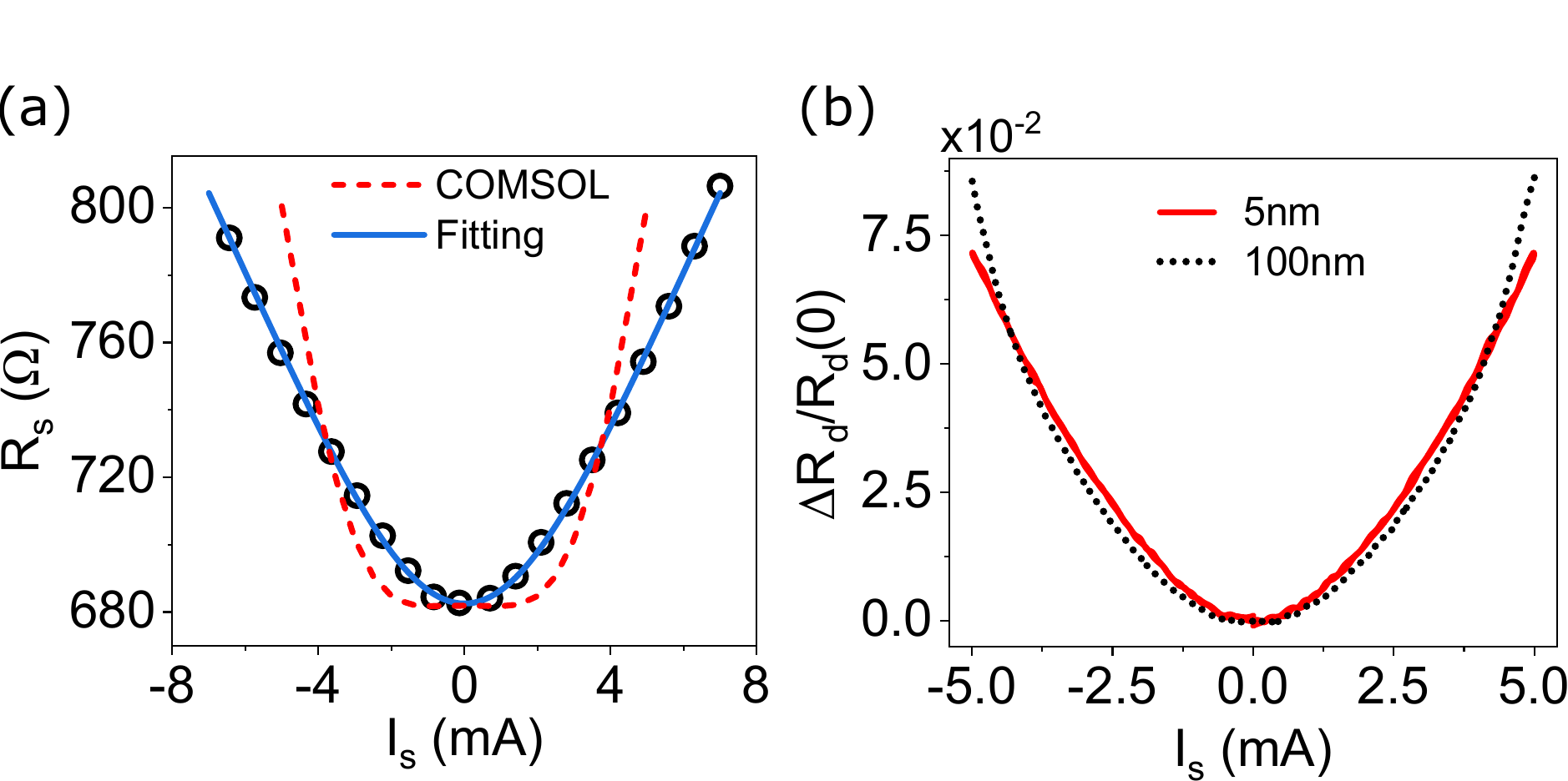}
	\caption{\label{fig:LT}
		(a) $R_s$ vs $I_s$ for $d=5$~nm at $T=7$~K. Symbols: data, dashed curve: COMSOL simulation, solid curve: fit with the function $\gamma(\alpha, \Delta I_s, I_s)$ defined in the text. (b) $\Delta R_d/R_d(0)$ vs $ I_s $ for $d=5$~nm (solid curve) and $100$~nm (dashed), at $T=7$~K. 
}	\end{figure}

Similar analysis predicts a quadratic $R_s(I_s)$, but with a coefficient that depends on the properties of the spacer. The resistivity of the Pt wires saturates below $\Theta_R$ [Fig.~\ref{fig:LT}(c)]. Therefore, for measurements performed at $\ll\Theta_R$, Joule heating should result in almost constant $R_s$ at small $I_s$, crossing over to a quadratic dependence at large $I_s$, as illustrated by the COMSOL simulation [dashed curve in Fig.~\ref{fig:LT}(a)]~\cite{kappaT}. 

The measured $R_s(I_s)$ at $T=7$~K [symbols in Fig.~\ref{fig:LT}(a)] is qualitatively different from this prediction. Instead, is can be well-fitted by $R_s(I)=R_s(0)+\gamma$, where $\gamma$ is a linear function of $|I_s|$ convolved with a Gaussian of width $\Delta I_s$, $\gamma(\alpha, \Delta I_s, I_s)=\alpha\int dI|I|e^{-(I-I_s)^2)/2\Delta I_s^2}/\sqrt{2\pi}\Delta I_s$  [solid curve in Fig.~\ref{fig:LT}(a)]~\cite{el-ph}. A similar result obtained in Ref.~\cite{chen2020observation} was interpreted as evidence for the non-equilibrium distribution of current-generated phonons that cannot be described by an effective temperature. We now outline this interpretation. According to the Drude-Sommerfeld model, the rate of electron scattering in the source wire is proportional to $I_s$. Assuming that one phonon is generated in each electron scattering event, and that phonons escape into the substrate before they can thermalize, one obtains from the kinetic balance relation a linear dependence of current-generated phonon population on current~\cite{chen2020observation}. Electron scattering on these phonons leads to a linear dependence $R_s(|I_s|)$. 

In this picture, phonon distribution must be non-thermal by energy conservation argument. The deposited electrical energy is quadratic in $I_s$, while the number of the generated phonons is proportional to $|I_s|$, and thus the average energy per phonon is also proportional to $|I_s|$. The same conclusion is obtained by considering the energy imparted by electric bias to each electron between the scattering events, which must be transferred to the phonon generated upon scattering. Smoothing of the weak singularity at $I_s=0$, accounted for by the Gaussian of width $\Delta I_s$, is explained by the reduced electron scattering cross-section on low-energy phonons. 

For thermalized phonon distribution below the Debye temperature $\Theta_D=240$~K of Pt, the average phonon energy $\left<\epsilon\right>$ is linear in $T$, as follows from the Debye integral $\left<\epsilon\right>\propto\int_{0}^{\infty}(e^x-1)^{-1}kTx^3dx/\int_{0}^{\infty}(e^x-1)^{-1}x^2dx\propto T$. Thus, for $T_s\propto I_s^2$, $\left<\epsilon\right>\propto I_{s}^2$. At $T>\Theta_D$, the average energy of thermalized phonons is independent of $T$, as follows from the Raleigh-Jeans law for the thermal mode population, $n(\epsilon)=kT/\epsilon$. One can conclude that linear dependence of $R_s$ on $|I_s|$ is inconsistent with thermal distribution of current-generated phonons.

Non-local measurements provide a test for this interpretation. If the distribution of phonons injected into the spacer is non-thermal, their inelastic scattering is expected to result in gradual thermalization. Thus, as the spacer thickness is increased, $R_d(I_s)$ may be expected to gradually transform from the linear dependence to the form expected for Joule heating. Indeed, $R_d(I_s)$ for $d=5$~nm is close to the linear dependence $R_s(I_s)$, while the dependence for $d=100$~nm is closer to parabolic [Fig.~\ref{fig:LT}(b)]. As a consequence,  at small $I_s$ the normalized $R_d$ is larger for $d=5$~nm than for $d=100$~nm, and smaller at large $I_s$, with a crossover at $|I_s|=4$~mA.

\begin{figure}
	\includegraphics[width=\columnwidth]{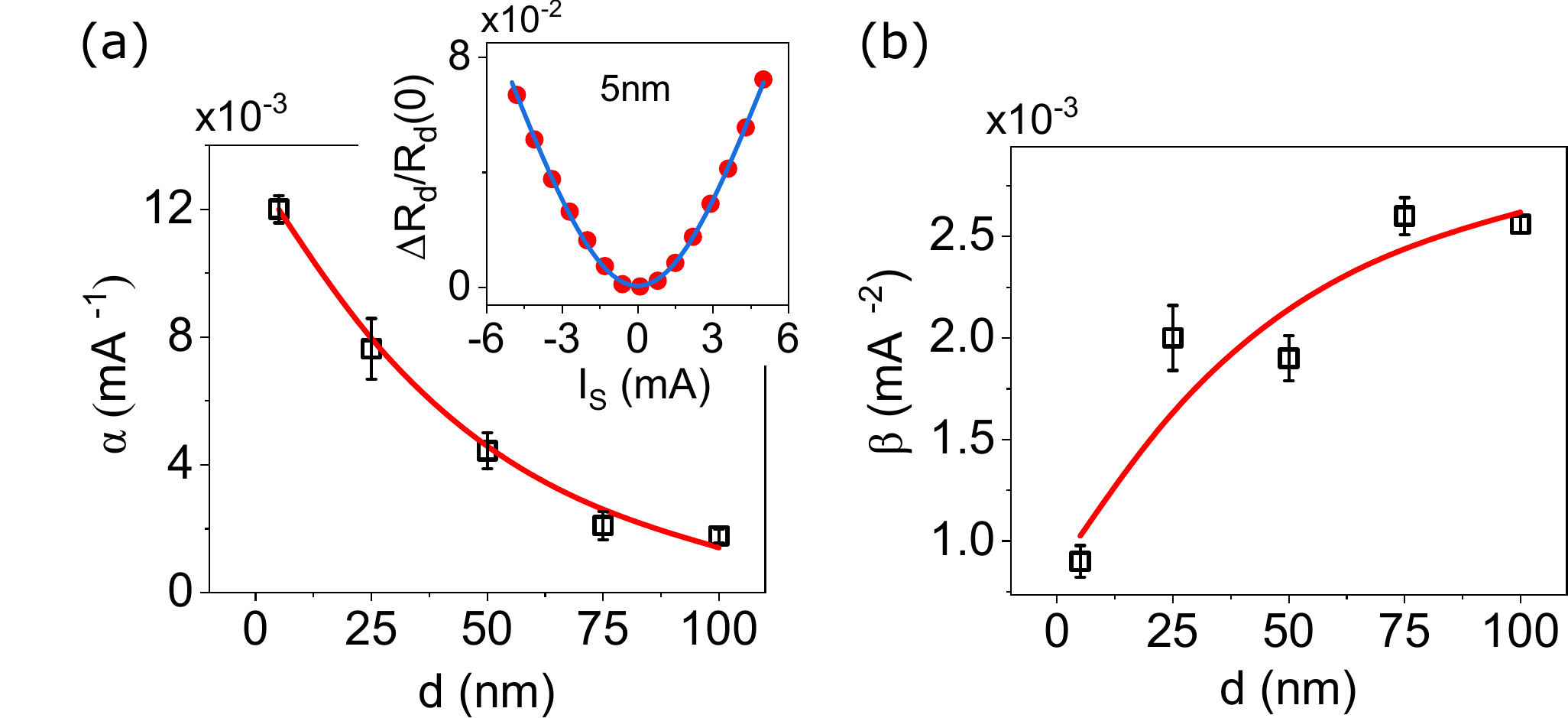}
	\caption{\label{fig:vsd} Dependence of the amplitudes $\alpha$ (a) and $\beta$ (b) of the linear and the quadratic contributions to $R_d(I_s)$ on $d$, determined from fits such as shown in the inset for $d=5$~nm, at $T=7$~K. Curves are fits with the exponential dependences $\alpha_0e^{-\ d/d_0}$ in (a), and $\beta_0-\beta_1e^{-d/d_0}$ in (b).}
\end{figure}

All the $R_d$ vs $I_s$ curves obtained at $T=7$~K were well-approximated by a sum of the function $\gamma(\alpha, \Delta I_s, I_s)$ used for fitting the local measurements, with the same value of $\Delta I_s$, and the quadratic function $\beta I_s^2$ [inset in Fig.~\ref{fig:vsd}(a)]. The first contribution describes the "primary" phonons generated in the source wire that diffuse to the detector without experiencing thermalizing inelastic scattering. The quadratic term was empirically found to provide a good fitting to the data. It reflects the presence of a ''secondary" group of phonons generated due to inelastic scattering of the "primary" phonons~\cite{optical}. 

The amplitude $\alpha$ of the linear contribution decreases, while the quadratic contribution increases with increasing $d$ [Fig.~\ref{fig:vsd}]. The dependence $\alpha(d)$ is well-approximated by the exponential decay with decay length $d_0=44\pm2$~nm [curve in Fig.~\ref{fig:vsd}(a)]. The variability of the quadratic amplitude $\beta$ is significantly larger. Nevertheless, it can be also fitted using the same exponential form describing its increase and saturation at large $d$ [curve in Fig.~\ref{fig:vsd}(b)]. These results are consistent with increasing effects of inelastic scattering that involves annihilation of the primary phonons and generation of the secondary phonons.

We note that $\beta(d)$ extrapolates to a finite value $\beta(0)=8\times 10^{-4}$~mA$^{-2}$, indicating that the secondary phonons are also generated at Pt/SiO$_2$ interfaces. We can estimate the probability $P_{el}$ that the primary phonon is transmitted elastically across the Pt/SiO$_2$ interface, by using the extrapolated value $\beta(\infty)=2.8\times10^{-3}$~mA$^{-2}$, which corresponds to all the primary phonons converted into secondary phonons in a thick spacer. Accounting for the two Pt/SiO$_2$ interfaces separating the source from the detector, we obtain $P_{el}^2=1-8/28$, i.e., $P_{el}=0.85$.

Based on the energy conservation arguments discussed above, the density $n_1$ of the primary phonons and their average quasiparticle energy $\epsilon_1$ are  proportional to $|I_s|$. The quadratic term in the dependence $R_d(I_s)$ indicates that the density $n_2$ of the secondary phonons is quadratic in $I_s$, and thus their average quasiparticle energy $\epsilon_2$ is independent of $I_s$. This implies that the secondary phonons generated at $T=7$~K are not thermalized, since the average energy of thermalized phonons at $T\ll \Theta_D$ would be proportional to $I_s^2$ [see above].

To analyze inelastic scattering that results in the generation of secondary phonons, we consider the continuity equations in the relaxation time approximation,
$\partial n_1/\partial t=-\nabla\cdot\mathbf{f}_1-n_1/\tau_{in}$ for the primary phonons, and $\partial n_2/\partial t=-\nabla\cdot\mathbf{f}_2+n_1\epsilon_1/\tau_{in}\epsilon_2$ for the secondary phonons. Here, $\mathbf{f}_{1,2}$ are the quasiparticle fluxes of the two phonon groups, and we used energy conservation to relate the secondary phonon generation to the annihilation rate $1/\tau_{in}$ of the primary phonons.

In the diffusive phonon transport approximation justified by the small phonon MFP $l_{el}\approx 5$~nm in amorphous SiO$_2$ at cryogenic temperatures~\cite{yang2017ballistic,kittel1949interpretation}, $\mathbf{f}_{1,2}=-D\nabla n_{1,2}$, where $D=v_{ph}l_{el}/3$ is the diffusion coefficient and $v_{ph}$ is the phonon group velocity, which can be approximated by the sound velocity $\approx 4.5$~km/s in amorphous SiO$_2$. Thus, we estimate $D\approx 7.5\times 10^{-6}$m$^2$/s.

The phonon distribution in the spacer depends only on the normal coordinate $z$, defined to be directed into the substrate, with the origin located in the SiO$_2$ spacer at the boundary with the source wire. Assuming that the only phonon source is at $z<0$, we obtain for the stationary state at $z>0$
\begin{equation}\label{eq:nvsz}
\begin{aligned}
n_1(z)&=n^{(0)}_1e^{-z/\sqrt{D \tau_{in}}},\\
n_2(z)&=n^{(0)}_1\frac{\epsilon_1}{\epsilon_2}+(n^{(0)}_2-n^{(0)}_1\frac{\epsilon_1}{\epsilon_2})e^{-z/\sqrt{D \tau_{in}}}
\end{aligned}
\end{equation}
where $n^{(0)}_1$ ($n^{(0)}_2$) is the primary (secondary) phonon density at $z=0$. These dependences are consistent with the observed exponential decay of $\alpha$ and the corresponding increase of $\beta$ [Fig.~\ref{fig:vsd}(b)], allowing us to estimate the inelastic scattering time $\tau_{in}=d_0^2/D=270$~ps. The inelastic scattering length, defined as the length of the phonon path between inelastic scattering events, is $l_{in}=\tau_{in}v_{ph}=1.2$~$\mu$m. Such a large value is promising for the possibility to transport non-thermalized phonons over significant distances, by utilizing materials with large phonon MFP.

\begin{figure}
	\includegraphics[width=\columnwidth]{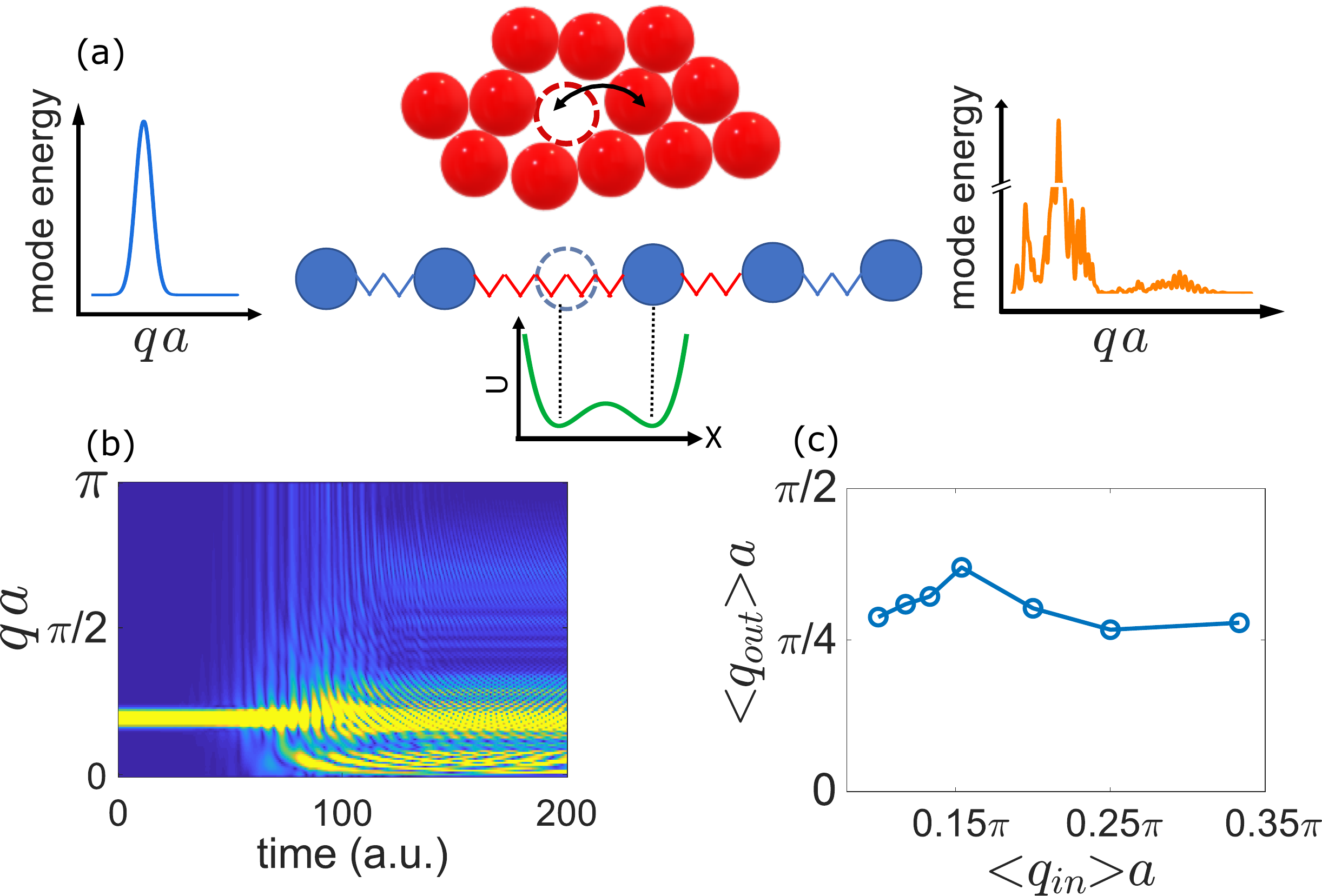}
	\caption{\label{fig:anharmonic}(a) Schematic of the 1d simulation of inelastic phonon scattering by a strongly anharmonic defect as a chain of masses connected by springs. Top inset: example of a strongly anharmonic defect associated with a bistable atomic position in amorphous solid. Bottom inset: potential vs displacement for the mass $n_0$ attached to the anharmonic springs. Left (right) inset: mode energy vs wavevector times the lattice constant for the phonon wavepacket before (after) scattering. (b) Wavevector distribution vs time during phonon scattering on the defect. (c) Average wavevector of the generated phonon modes vs the incident phonon wavevector, obtained by cutting out twice the Guassian width around the wavepacket center after scattering.
}
\end{figure}

We now analyze the mechanisms of secondary phonon generation. Inelastic scattering is usually described in terms of three- and four-phonon processes~\cite{ziman2001electrons,PhysRevB.102.195412,Liao2021}. In these processes governed by quasiparticle energy and momentum conservation, the average energies of the generated phonons linearly scale with the energies of the annihilated phonons, which is inconsistent with our results. A cascade of three-phonon processes could  result in effective thermalization. However, this cannot explain our observation of similar behaviors over a wide range of spacer thicknesses. Damping of long-wavelength acoustic waves can be described by the diffusive Akhiezer mechanism associated with the strain-induced modulation of phonon spectrum~\cite{Akhiezer1939}. However, this quasi-adiabatic mechanism is not applicable to high-energy nonequilibrium phonons generated at large electric bias.

Since inelastic scattering requires anharmonicity, but weak anharmonicity responsible for the few-phonon processes cannot account for our observations, we conclude that secondary phonons are likely generated due to strongly anharmonic dynamics. Such dynamics may be related to the bosonic peak ubiquitous in the spectra of amorphous materials~\cite{Zhang2021,Ciamarra2021,Zaccone2021} and to the problem of phonon glass~\cite{PhysRevB.48.12589,Allen1999}, and may be associated with quasi-localized nonlinear defect states such as bistable atomic configurations [top inset in Fig.~\ref{fig:anharmonic}(a)], as well as interstitial impurities and incoherent interfaces~\cite{pohl2002low,molina2021decoupling}.

To model phonon scattering by a bistable defect, we consider a wavepacket propagating along a 1d chain of masses $n=1..300$ connected to their neighbors by springs [Fig.~\ref{fig:anharmonic}(a)]. Masses $n=1$ and $300$ are also connected to avoid artifacts from boundary reflections. All the springs are linear with the same spring constant, except the two springs connected to the mass $n_0=150$ are described by the double-well potential energy
$U(x)=-k_1x^2/2+k_3x^4/4$, resulting in bistable equilibrium of mass $n_0$  [bottom inset in Fig.~\ref{fig:anharmonic}(a)].

Dynamics spanning both potential wells cannot be described in terms of the perturbative anharmonic expansion because of the saddle point of the potential at $x=0$. Scattering of the wavepacket on the defect results in the generation of a broad range of modes throughout the entire Brillouin zone, Fig.~\ref{fig:anharmonic}(b), instead of the usual harmonics expected for weak anharmonicity~\cite{Landau1976Mechanics}. Figure.~\ref{fig:anharmonic}(c) shows that the average wavevector of the generated modes remains almost constant when the center wavevector of the incident wavepacket is varied by more than a factor of $5$, consistent with our experimental observations. This result indicates a breakdown of the perturbative picture underlying the concept of quasiparticles [phonons]~\cite{arXiv:2106.08459}, enabling nonresonant scattering not constrained by the usual quasiparticle momentum and energy conservation laws.

In summary, we studied non-equilibrium phonons generated by current in a ''source" Pt nanowire, by measuring the resistance of a ''sensing" nanowire insulated from the source by a SiO$_2$ spacer. At cryogenic temperatures, the dependence of resistance on current can be precisely fitted by the sum of a linear and a quadratic function. The linear contribution exponentially decreases, while the quadratic one similarly increases with increasing spacer thickness. We interpret these results as evidence for a highly non-thermal distribution of current-generated phonons, which relax by inelastic scattering mediated by strongly anharmonic defects that alleviate the constraints imposed by the quasiparticle momentum and energy conservation. Our results suggest a new route for the characterization of nonequilibrium current-generated phonons, and for optimizing Joule heat management in electronic nanodevices.

We thank Vasili Perebeinos for a helpful discussion. This work was supported by the U.S. Department of Energy, Basic Energy Sciences, Award \# DE-SC0018976.

\bibliography{ph_nonlocal}

\begin{thebibliography}{45}
\bibitem [{tem()}]{temperature}%
  \BibitemOpen
  \href@noop {} {}\bibinfo {note} {The adjective ''effective" reflects the fact
  that temperature is a rigorous characteristic only in true
  equilibrium}\BibitemShut {NoStop}%
\bibitem [{hig()}]{highT}%
  \BibitemOpen
  \href@noop {} {}\bibinfo {note} {Room-temperature measurements for $d<25$~nm
  were unreliable because of electron activation across spacer}\BibitemShut
  {NoStop}%
\bibitem [{kap()}]{kappaT}%
  \BibitemOpen
  \href@noop {} {}\bibinfo {note} {The simulation accounts for the temperature
  dependences of thermal conductivity of the spacer and the substrate, which do
  not significantly affect this argument}\BibitemShut {NoStop}%
\bibitem [{el-()}]{el-ph}%
  \BibitemOpen
  \href@noop {} {}\bibinfo {note} {Phonon generation due to electron-phonon
  scattering can be neglected, since resistance is dominated by scattering at
  the boundaries and defects even at large bias}\BibitemShut {NoStop}%
\end{thebibliography}
\bibliographystyle{apsrev4-1}
	
\end{document}